# Fractional Order Fuzzy Control of Hybrid Power System with Renewable Generation Using Chaotic PSO

Indranil Pan[a] and Saptarshi Das[b,c]

a) Centre for Energy Studies, Indian Institute of Technology Delhi, Hauz Khas, New Delhi 110016, India.

b) School of Electronics and Computer Science, University of Southampton, Southampton SO17 1BJ, United Kingdom.

c) Department of Power Engineering, Jadavpur University, Salt Lake Campus, LB-8, Sector 3, Kolkata-700098, India.

**Authors' Emails:**

indranil.jj@student.iitd.ac.in (I. Pan)

saptarshi@pe.jusl.ac.in, s.das@soton.ac.uk (S. Das*)

**Phone:** +44- 7448572598

## Abstract:

This paper investigates the operation of a hybrid power system through a novel fuzzy control scheme. The hybrid power system employs various autonomous generation systems like wind turbine, solar photovoltaic, diesel engine, fuel-cell, aqua electrolyzer etc. Other energy storage devices like the battery, flywheel and ultra-capacitor are also present in the network. A novel fractional order (FO) fuzzy control scheme is employed and its parameters are tuned with a particle swarm optimization (PSO) algorithm augmented with two chaotic maps for achieving an improved performance. This FO fuzzy controller shows better performance over the classical PID, and the integer order fuzzy PID controller in both linear and nonlinear operating regimes. The FO fuzzy controller also shows stronger robustness properties against system parameter variation and rate constraint nonlinearity, than that with the other controller structures. The robustness is a highly desirable property in such a scenario since many components of the hybrid power system may be switched on/off or may run at lower/higher power output, at different time instants.

## 1. Introduction

The increase in energy demand coupled with the rising concerns of global warming, has necessitated the integration of renewable energy technologies like wind and solar energy into the power grid. This has given rise to hybrid distributed energy generation and storage systems [1]. The generation from the wind and solar power plants are stochastic in nature and depend on the weather conditions at any particular time instant. This might result in situations where the electrical load is higher than the generation. Energy storage devices like batteries, flywheels or ultra-capacitors might be coupled with such systems to mitigate this unbalance.

They also improve the power quality and decrease the fluctuations in grid frequency [2]. If there is surplus power available from these renewable sources over the demanded load, these storage devices store them for a short period of time and later release them to the grid when the demand load is higher than the generation. For these actions to be performed properly there needs to be a control strategy which coordinates these activities.

Control systems based on fractional calculus [3] is gaining increasing interest in the research community due to its additional flexibility and superior design performance [4][5]. Fractional calculus has spurred recent interest in signal processing [6] and computational intelligence techniques have also been integrated in the design of fractional order control systems [7]. These fractional order intelligent control systems are finding wide applications in process control [8], [9], nuclear reactor control [10], chaos synchronization [11] etc. among many others. Fractional calculus has also been integrated with fuzzy logic [12][13] and PSO [14], [15] to enhance their performance. Also, computational intelligence based design for fractional order control systems have been found expedient in different power system applications like automatic voltage regulator [16–18], two area load frequency control [19], microgrid frequency control [20] etc. Motivated by the success of such diverse applications of computational intelligence based fractional order control systems, a fractional order fuzzy control scheme is explored in this paper for the case of hybrid power systems. Other approaches towards designing control strategy for these kind of systems include the standard PID controller [21], genetic algorithm based PI/PID controller [22], robust $H_\infty$ controller [23][24] etc.

In this paper, a comparison has been reported between standard PID and fuzzy PID controller to show the advantage of the proposed scheme. Due to the presence of stochastic renewable energy generation components like wind and solar power, there is a continuous variation in grid frequency. This affects the power quality which needs to be kept within limits so that the downstream connected electrical loads do not malfunction. For this purpose a controller is introduced in the loop, which sends a signal to the energy storage systems to absorb/release additional/deficit power from/into the grid respectively. The controller also sends a command to the diesel engine to release high bursts of power into the grid to meet short term load demands. The fractional order fuzzy PID controller [25] is employed for this purpose and is compared with performances achieved by PID and fuzzy PID controller. The schematic of the hybrid power system along with the controller is illustrated in Figure 1. Another advantage of our proposed control scheme over that reported in [18]-[19] is that only one centralized controller structure is required for the overall hybrid power system. This removes the necessity of having one controller for each of the power storage units in the feedback path like that reported in [21][22] and eliminates the need for effective tuning of each of the controllers simultaneously, which is cumbersome in practice. The proposed scheme, therefore, reduces cost, additional wiring, maintenance and also the necessity of tuning each of the controllers separately, avoiding any possible performance deterioration due to loop interactions.

The evolutionary and swarm algorithm works well over the classical gradient based methods especially in noisy [26] and dynamic environments [27], [28], [29] which are commonly encountered in control system design to handle stochastic fluctuations. In this paper, a chaotic PSO algorithm has been used for determining the controller parameters by optimizing a time domain performance metric (which is noisy) due to the presence of stochastic fluctuations in the wind power generation, solar power generation and the load profile. Other new search and optimization algorithms like the gravitational search algorithm [30] and cat swarm algorithm [31] might also be used to find the controller parameters.

The rest of the paper is organized as follows. Section 2 describes the various components of the hybrid energy system. Section 3 briefly introduces the FO fuzzy controller which keeps the frequency deviation of the power system within allowable range. Section 4 explains two-chaotic map adapted versions of PSO algorithm, along with the objective function for optimization. Section 5 presents the comparison amongst the performances of three controller structures and also their robustness against system parameter variation. The effect of the rate constraint nonlinearity in the feedback elements are explored next in section 6, followed by discussion on the heuristics and implementation issues in Section 7. The paper ends with the conclusions in Section 8, followed by the references.

## 2. Description of the hybrid power system with renewable generation

The schematic of the hybrid power system using different modes of energy generation and storage is illustrated in Figure 1 with its different components described in Table 1.

### 2.1. Models of various generation subsystems

For small signal analysis, the transfer functions of the WTG, STPG, FC and DEG can be modeled by first order transfer functions (1)-(4) with the associated gain and time constants given in Table 1 [32][21][22].

$$G_{WTG}(s) = K_{WTG}/(1+sT_{WTG}) = \Delta P_{WTG}/\Delta P_W \quad (1)$$

$$G_{STPG}(s) = \left(K_S/(T_S s+1)\right) \cdot \left(K_T/(T_T s+1)\right) = \Delta P_{STPG}/\Delta P_{sol} \quad (2)$$

$$G_{FC_k}(s) = K_{FC}/(1+sT_{FC}) = \Delta P_{FC_k}/\Delta P_{AE}, \; k=1,2 \quad (3)$$

$$G_{DEG}(s) = K_{DEG}/(1+sT_{DEG}) = \Delta P_{DEG}/\Delta u \quad (4)$$

TABLE 1: NOMINAL PARAMETERS OF THE COMPONENTS OF HYBRID POWER SYSTEM

| Component | Gain ($K$) | Time constant ($T$) |
|---|---|---|
| Wind turbine generator (WTG) | $K_{WTG}$=1 | $T_{WTG}$=1.5 |
| Aqua Electrolyzer (AE) | $K_{AE}$=0.002 | $T_{AE}$=0.5 |
| Fuel Cell (FC) | $K_{FC}$=0.01 | $T_{FC}$=4 |
| Flywheel energy storage system (FESS) | $K_{FESS}$=-0.01 | $T_{FESS}$=0.1 |
| Battery energy storage system (BESS) | $K_{BESS}$=-0.003 | $T_{BESS}$=0.1 |
| Ultracapacitor (UC) | $K_{UC}$=-0.7 | $T_{UC}$=0.9 |
| Diesel engine generator (DEG) | $K_{DEG}$=0.003 | $T_{DEG}$=2 |
| Solar Thermal Power Generation (STPG) | $K_S$=1.8, $K_T$=1 | $T_S$=1.8, $T_T$=0.3 |

### 2.2. Model of the aqua electrolyzer

The aqua-electrolyzer produces hydrogen for the fuel cell using a part of the power generated from the renewable energy sources like wind and/or solar. The dynamics of the AE for small signal analysis can be represented by the transfer function (5) [32] where it uses $(1-K_n)$ fraction of the total power of WTG and STPG to produce hydrogen which is again used by two FCs to produce power and feed it to the grid.

$$G_{AE}(s) = K_{AE}/(1+sT_{AE}) = \Delta P_{AE}/((\Delta P_{WTG} + \Delta P_{STPG})(1-K_n)) \quad (5)$$

where, $K_n = P_t/(P_{WTG} + P_{STPG}), K_n = 0.6$ (6)

### 2.3. Model of different energy storage systems

In the hybrid energy system of Figure 1, the FESS, BESS and the UC are connected in the feedback loop and are actuated by the signal from the FO fuzzy controller. These absorb or release energy from or to the grid if there is a surplus or deficit amount of power respectively. Their corresponding transfer functions can be represented as (7)-(9) [32] where $\Delta u$ is the incremental control action by the centralized controller employed in feedback path, to reduce the grid frequency oscillation $\Delta f$.

$$G_{FESS}(s) = K_{FESS}/(1+sT_{FESS}) = \Delta P_{FESS}/\Delta u \quad (7)$$

$$G_{BESS}(s) = K_{BESS}/(1+sT_{BESS}) = \Delta P_{BESS}/\Delta u \quad (8)$$

$$G_{UC}(s) = K_{UC}/(T_{UC}s+1) = \Delta P_{UC}/\Delta u \quad (9)$$

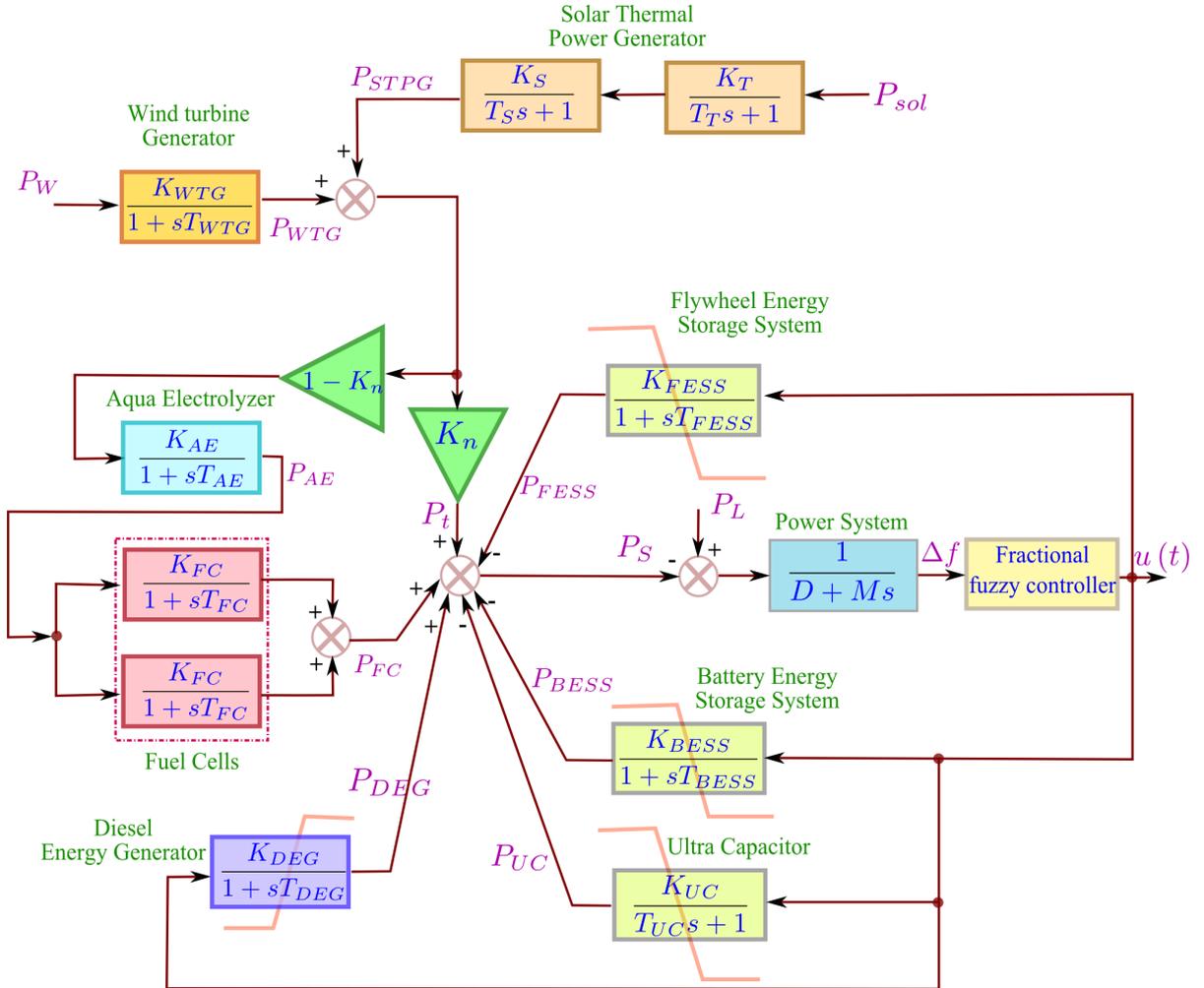

Figure 1: Schematic of the hybrid power system with rate constraint nonlinearity in energy storage/generation elements.

### 2.4. Power system model using grid frequency deviation

The power system model can be represented as (10).

$$G_{sys}(s) = \Delta f/\Delta P_e = 1/(D+Ms) \quad (10)$$

ISA Transactions

where, $M$ and $D$ are the equivalent inertia constant and damping constant of the hybrid power system [1] and their typical values are considered as 0.4 and 0.03 respectively for the present simulation study.

## *2.5.    Stochastic model of the renewable energy: wind and solar power generation and the demand load*

The wind generation, solar generation and load demand, is modeled in a general template considering both large deterministic drift and small stochastic fluctuations [22]. The models result in a mean value and stochastic fluctuations about the mean generated or demand power at each time instant. Additionally there is a sudden shift in the mean value at some point in time to indicate a greater variation in these parameters. The general template for these is chosen as (11)

$$P = \left(\left(\phi\eta\sqrt{\beta}\left(1-G(s)\right)+\beta\right)\delta/\beta\right)\Gamma = \xi \cdot \Gamma \qquad (11)$$

where, $P$ represents the power output of the solar, wind or the load model, $\phi$ is the stochastic component of the power, $\beta$ contributes to the mean value of the power, $G(s)$ is a low pass transfer function, $\{\eta, \delta\}$ are constants in order to normalize the generated or demand powers $\xi$ to match the per unit (p.u.) level, $\Gamma$ is a time dependent switching signal with a gain which dictates the sudden fluctuation in mean value for the power output.

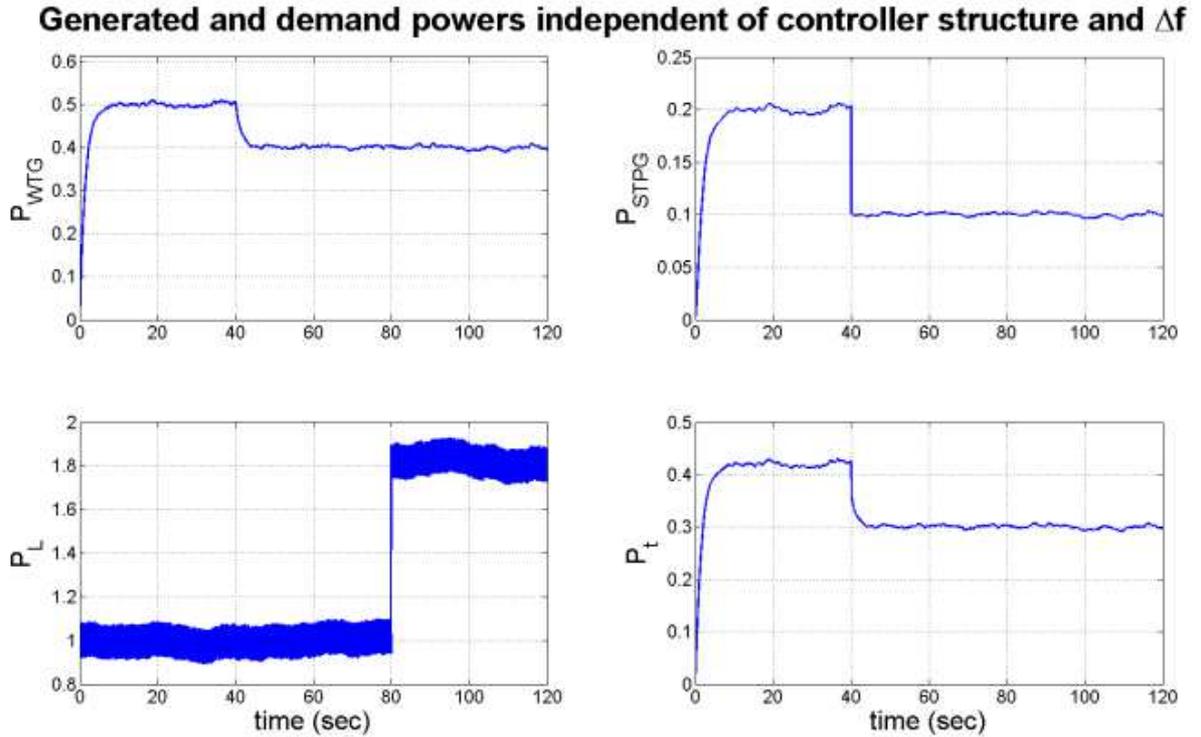

Figure 2: A single realization of the renewable power generation and demand load.

For the wind power generation, the parameters of (11) are $\phi \sim \mathbb{U}(-1,1)$, $\eta = 0.8$, $\beta = 10$, $G(s) = 1/(10^4 s + 1)$, $\delta = 1$ and $\Gamma = 0.5H(t) - 0.1H(t-40)$, where $H(t)$ is the Heaviside step function. For the solar power generation, the parameters of (11) are $\phi \sim \mathbb{U}(-1,1)$, $\eta = 0.7$, $\beta = 2$, $G(s) = 1/(10^4 s + 1)$, $\delta = 0.1$ and $\Gamma = 1.1111 H(t) - 0.5555 H(t-40)$. For the demand load, the parameters of (11) are $\phi \sim \mathbb{U}(-1,1)$, $G(s) = (300/(300s+1)) - (1/(1800s+1))$, $\eta = 0.8$, $\beta = 100$, $\delta = 1$, and $\Gamma = H(t) + (0.8/\xi) H(t-80)$.

A single realization of the stochastic components *viz.* generated powers ($P_W, P_{sol}$), demand ($P_L$) and also the net generated power to the grid ($P_t$) are shown in Figure 2. These are independent of the controller structure present in the feedback path. It can be seen for all the cases that there is a stochastic component superimposed on a base value and there are sudden jumps of the base value at arbitrary instants of time to indicate a sudden large change in the power at different time instants (40 sec and 80 sec in this case). The expression (11) with parameters, mentioned above have been used to calculate the net power generated by the wind and solar units i.e. $\{P_W, P_{sol}\}$ in (1) and (2) respectively.

## 3. Fractional order fuzzy controller

### 3.1. Basics of fractional calculus

Fractional calculus is an extension of the $n^{th}$ order successive differentiation and integration of an arbitrary function having the order as any real value. There are three main definitions of fractional calculus, the Grünwald-Letnikov (GL), Riemann-Liouville (RL) and Caputo definitions [3]. The Caputo definition is widely used in fractional order control system design problems [3–7]. According to Caputo's definition, the $\alpha^{th}$ order differ-integral of a function $f(t)$ with respect to time $t$ is given by (12).

$$D^\alpha f(t) = \frac{1}{\Gamma(m-\alpha)} \int_0^t \frac{D^m f(\tau)}{(t-\tau)^{\alpha+1-m}} d\tau, \quad (12)$$

$$\alpha \in \mathbb{R}^+, m \in \mathbb{Z}^+, m-1 \leq \alpha < m$$

### 3.2. Hybridization of fuzzy PID and fractional order control

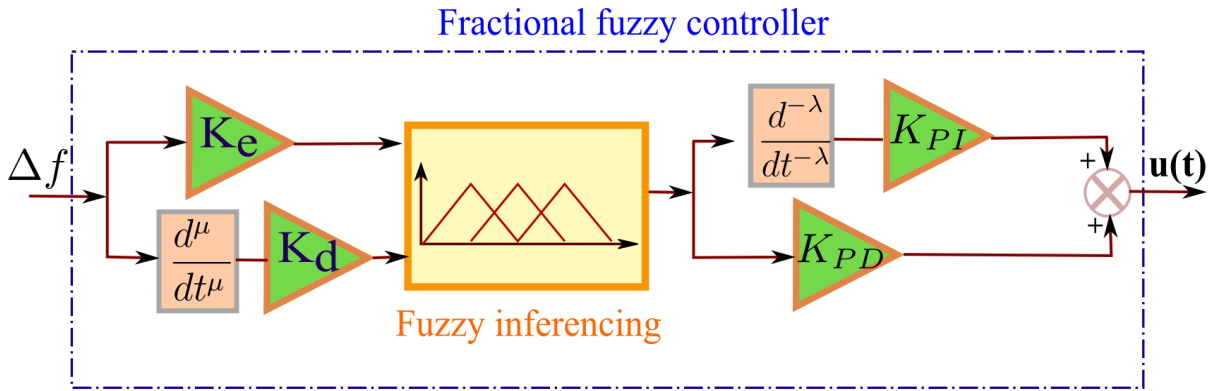

Figure 3: Schematic of the fractional order fuzzy PID controller.

The FO fuzzy PID controller has been introduced by Das *et al.* in [25] with $\{K_e, K_d\}$ and $\{K_{PI}, K_{PD}\}$ being its input and output scaling factors (SFs) respectively and has been shown to give good results for process control applications [8], [10], [25], [26]. The schematic diagram of the fuzzy FOPID controller is shown in Figure 3. Also, the rule base considered for the fuzzy controller is depicted in Figure 4 and the corresponding membership functions in Figure 5. The fuzzy linguistic variables NL, NM, NS, ZR, PS, PM, PL represent Negative Large, Negative Medium, Negative Small, Zero, Positive Small, Positive Medium and Positive Large respectively. The crisp output of the Fuzzy Logic Controller is determined by using center of gravity method of defuzzification. The FO fuzzy controller SFs and integro-differential orders $\{K_e, K_d, K_{PI}, K_{PD}, \lambda, \mu\}$ are tuned using PSO for a fixed rule base and membership function type.

| $\dfrac{d^\mu e}{dt^\mu}$ \ $e$ | NL | NM | NS | ZR | PS | PM | PL |
|---|---|---|---|---|---|---|---|
| PL | ZR | PS | PM | PL | PL | PL | PL |
| PM | NS | ZR | PS | PM | PL | PL | PL |
| PS | NM | NS | ZR | PS | PM | PL | PL |
| ZR | NL | NM | NS | ZR | PS | PM | PL |
| NS | NL | NL | NM | NS | ZR | PS | PM |
| NM | NL | NL | NL | NM | NS | ZR | PS |
| NL | NL | NL | NL | NL | NM | NS | ZR |

Figure 4: Rule base for error, fractional rate of error and FLC output.

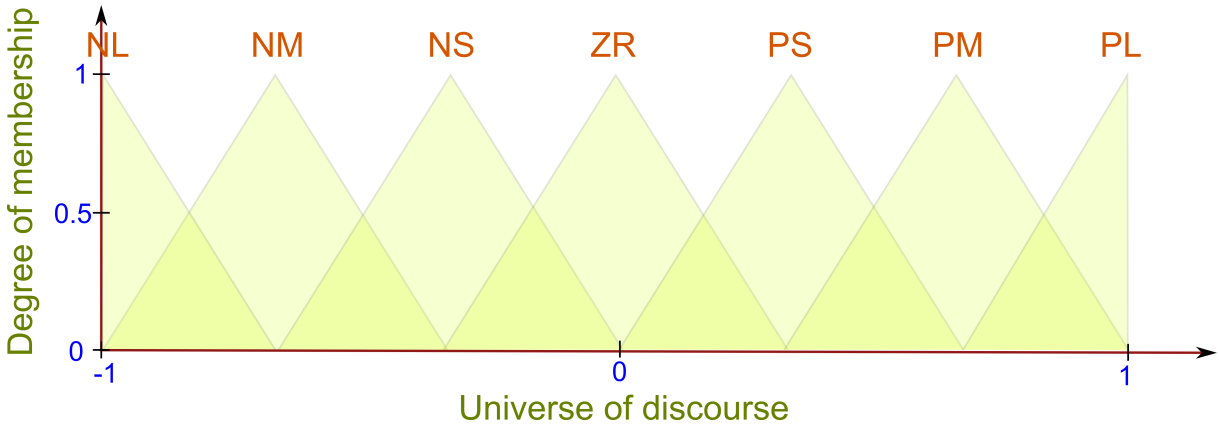

Figure 5: Rule base for error, fractional rate of error and FLC output.

Various continuous and discrete time rational approximation methods exist for fractional order elements [4][3][5] which can be seen as the heart of the proposed fuzzy logic based fractional order controller. In the present paper, each guess value of the fractional order differ-integrals $\{\lambda, \mu\}$ within the optimization process is continuously rationalized with Oustaloup's 5$^{th}$ order rational approximation. The FO differ-integrators are basically infinite dimensional linear filters. However, band-limited realizations of fractional order controllers are necessary for practical implementation. In the present study, each fractional order element has been rationalized with Oustaloup's recursive filter given by the equations (13)-(14). If it be assumed that the expected fitting range or frequency range of controller operation is $(\omega_b, \omega_h)$, then the higher order filter which approximates the FO element $s^\alpha$ can be written as (13).

$$G_f(s) = s^\alpha \approx K \prod_{k=-N}^{N} \left((s+\omega_k')/(s+\omega_k)\right) \tag{13}$$

Its poles, zeros, and gain of the filter can be evaluated as (14).

$$\omega_k = \omega_b \left(\omega_h/\omega_b\right)^{\frac{k+N+(1+\alpha)/2}{2N+1}}, \omega_k' = \omega_b \left(\omega_h/\omega_b\right)^{\frac{k+N+(1-\alpha)/2}{2N+1}}, K = \omega_h^\alpha \tag{14}$$

In equations (13)-(14), $\alpha$ is the order of the differ-integration and $(2N+1)$ is the order of the realized analog filter. The present study considers a 5$^{th}$ order Oustaloup's approximation for all the FO elements within the frequency range of $\omega \in \{10^{-2}, 10^2\}$ rad/sec.

## 4. Control objectives and optimization based tuning of the fuzzy FOPID controller parameters

### 4.1. Optimization strategy for dynamically changing objective functions

The objective function ($J$) for optimization has been considered as an integral performance index over the simulation period of $T_{max} = 120$ sec, using the weighted sum of squared frequency deviation and the deviation of control signal $u$ from its expected steady state value $u_{ss}$ as given by (15).

$$J = \int_0^{T_{max}} \left(w_1 (\Delta f)^2 + w_2 (u - u_{ss})^2\right) dt \tag{15}$$

In (15) the first term represents the Integral of Squared Error (ISE) of grid frequency deviation and the second term is known as the Integral of Squared Deviation of Controller Output (ISDCO) as studied in [8], [20] for a disturbance rejection task of the controller, since it is placed in the feedback path. In (15), the weights $w_1, w_2$ govern the relative importance of each term in the objective function. They are taken as $w_1 = w_2 = 1$ to give equal importance to both. Here, the steady state control signal $u_{ss}$ changes after each switching in the generation and load as also studied in [32][20]. For the present simulation study, $T_{max} = 120$ sec and $u_{ss} = 0.81H(t) + 0.17H(t-40) + 1.12H(t-80)$.

Due to the presence of stochastic terms in generation and load as shown in Figure 2, the optimization for controller tuning essentially deals with locating the expected minima of a dynamic (time-varying) objective function [10], [27], [28], [33]. The PSO variants are employed to identify the expected minima of the multiple realizations of the objective function (15) that slightly changes its shape in each realization (ensembles) and fluctuates around the respective steady values during the three time intervals in $u_{ss}$ [33], [27]. Also, the objective function (15) is formulated in such a way that beside the frequency oscillation ($\Delta f$), the control signal variations to different actuators are also minimum to avoid possible mechanical shock/stress in those elements. This helps to limit the requirement of increased capacity for the battery and ultra-capacitor, reduces flywheel jerk and diesel consumption, making the overall hybrid power system more cost-effective.

## *4.2. Chaotic map adapted particle swarm optimization*

The PSO algorithm tries to optimize an objective function $f(x)$ with respect to the design variable $x \in \mathbb{R}^n$. It is expressed as (16).

$$\underset{x \in \mathbb{R}^n}{\text{minimize}}\ f(x) \tag{16}$$

where, the objective function $f : \mathbb{R}^n \to \mathbb{R}$ and the $n$-dimensional search space $G \in \mathbb{R}^n$ is pre-specified by the user. The PSO algorithm consists of a swarm of particles $x_i\ \forall\ i \in \{1, 2, ..., n_p\}$. The maximum number of particles $n_p$ is specified by the user. The particles $x_i$ search for an optimal solution $x' \in \mathbb{R}^n$ of (16). The position of the $i^{th}$ particle is denoted by $x_i := (x_{i,1}, x_{i,2}, ..., x_{i,n})^T \in \mathbb{R}^n$ and the velocity is denoted by $v_i := (v_{i,1}, v_{i,2}, ..., v_{i,n})^T \in \mathbb{R}^n$, where $i \in \{1, 2, ..., n_p\}$. The position and velocity of the $i^{th}$ particle $x_i \in \mathbb{R}^n$ is updated in each iteration, based on equations (17)-(18) for $k \in \mathbb{Z}^+$ which indicates the iteration number.

$$x_i^{k+1} = x_i^k + v_i^{k+1} \tag{17}$$

$$v_i^{k+1} = \alpha v_i^k + \beta_1 \theta_{1,i}^k \left(x_i^{best,k} - x_i^k\right) + \beta_2 \theta_{2,i}^k \left(x_{swarm}^{best,k} - x_i^k\right) \tag{18}$$

where, $\alpha$ is the inertia factor, $\beta_1$ is the cognitive learning rate and $\beta_2$ is the social learning rate and is pre-specified by the user. These influence the exploration and exploitation properties of the particles and must be properly chosen for faster convergence. $\theta_{1,i}^k$ and $\theta_{2,i}^k$ represent random numbers uniformly distributed in the interval $[0,1]$. $x_i^{best,k}$ in (19) refers to the previously obtained best position of the $i^{th}$ particle and $x_{swarm}^{best,k}$ denotes the best position of the swarm at the current iteration $k$. This is expressed as (20).

$$x_i^{best,k} := \arg\min_{x_i^j} \left\{f(x_i^j), 0 \le j \le k\right\} \tag{19}$$

$$x_{swarm}^{best,k} := \arg\min_{x_i^k} \left\{f(x_i^k), \forall i\right\} \tag{20}$$

The pseudo-code for the PSO algorithm can be summarized as follows:

**[*Step 1*]:** Initialize $n_p$ particles randomly distributed in the search space $G \in \mathbb{R}^n$ and calculate the objective function values for each particle $x_i\ \forall\ i \in \{1, 2, ..., n_p\}$. Set, $k = 0$. Determine $x_i^{best,0}$ and $x_{swarm}^{best,0}$.

**[*Step 2*]:** If the criteria for termination is satisfied, the algorithm terminates with the solution $x' := \arg\min_{x_i^j} \left\{f(x_i^j), \forall i, j\right\}$. Otherwise go to [Step 3].

**[*Step 3*]:** Use (17) and (18) to update the position and velocity of the particles and evaluate the corresponding objective functions at each position. Set, $k = k+1$. Determine $x_i^{best,k}$ and $x_{swarm}^{best,k}$ and go to [Step 2].

The termination criterion is set as the user specified maximum number of iterations $k_{max}$. The population is taken as 30 and the number of generations as 300. The inertia weight $\alpha$ is linearly decreased over the iterations from 0.9 to 0.1. The values of $\beta_1$ and $\beta_2$ are chosen as 0.5 and 1 respectively.

In [34], [35], it is shown that incorporating a chaotic map for the random number generation instead of the conventional random number generators (RNG), increases the efficiency of the algorithm and introduces diversity in the solutions. In the present study, a Henon map and a logistic map is coupled with the PSO algorithm for improved performance [18], [19]. The Henon map is a two dimensional discrete time dynamical system that exhibits chaotic behavior. Given a point with co-ordinates $\{x_n, y_n\}$, the Henon map transforms it to a new point $\{x_{n+1}, y_{n+1}\}$ using the following set of equations in (21).

$$x_{n+1} = y_n + 1 - ax_n^2,$$
$$y_{n+1} = bx_n \tag{21}$$

The map is chaotic for the parameters $a = 1.4$ and $b = 0.3$. It is actually a simplified model of the Poincare section of the Lorenz system. The initial values of all the variables are zero. The output $y_{n+1}$ varies in the range $[-0.3854, 0.3819]$. Since the Henon map is used here as a replacement for the RNG, it has to produce a number in the range $[0,1]$. Hence, the output is scaled in the range $[0,1]$ as also done in [36], [37].

The one dimensional chaotic Logistic map is given in (22).

$$x_{n+1} = ax_{n+1}(1 - x_n) \tag{22}$$

The initial condition of the map in (22) has been chosen to be $x_0 = 0.2027$ and the parameter $a = 4$ has been taken similar to that in [36], [37]. The performances of standard global optimizers have been improved to a higher extent using several chaotic maps reported as in [38], [39] and have also been applied in various power or energy system applications e.g. economic load dispatch [34], [35], automatic voltage regulator design [16], [18] and two area load frequency control [19] etc.

## 5. Results and discussions

### 5.1. *Performance of the controller in nominal condition of the hybrid power system*

The hybrid power system is simulated with different controller structures and optimized with the chaotic map adapted PSO algorithm as in section 4, while all the components in Figure 1 are considered to be working within the linear operating regime. The total simulation time is considered as 120 seconds and the coupled ordinary differential equations (comprising of the system components and controller) along with the stochastic forcing terms, representing the whole hybrid power system in Figure 1, are numerically integrated with a fixed step size of 0.01 sec using the 3$^{rd}$ order accurate Bogacki-Shampine formula. It has been found that the PID controllers tuned with different chaotic maps result in the same minima of the cost function as $J_{min} = 4.51$ with corresponding gains as $K_p = 2.04$, $K_i = 0.64$, $K_d = 0.61$, although the individual convergence characteristics are different for three PSO variants. Similar analysis for the fuzzy PID and fuzzy FOPID controllers are shown in Table 2 reporting the best controller parameters, corresponding to the minimum $J_{min}$ across multiple runs. The convergence characteristics for different PSO algorithms for each of the controllers are shown in Figure 6.

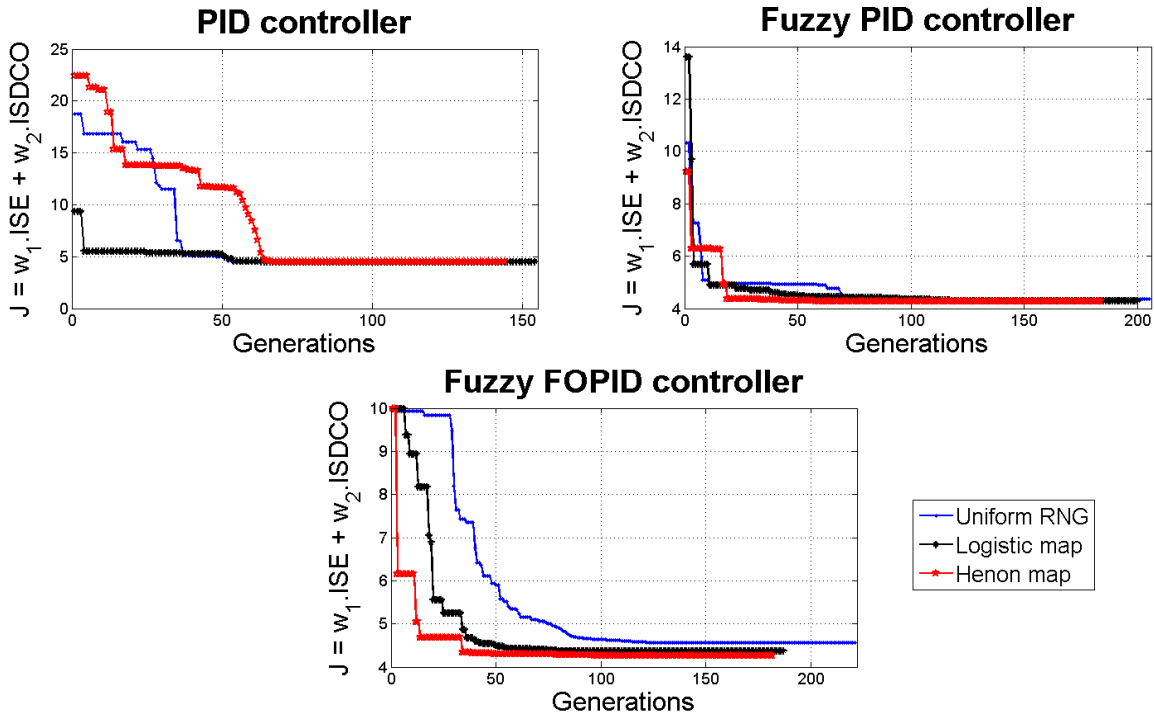

Figure 6: Convergence characteristics of the PSO algorithm with the uniform RNG and chaotic Logistic/Henon map.

TABLE 2: PSO BASED TUNING OF FUZZY PID AND FUZZY FOPID CONTROLLERS

| Controller structure | Random number generator | Controller parameters | | | | | | |
|---|---|---|---|---|---|---|---|---|
| | | $J_{min}$ | $K_e$ | $K_d$ | $K_{PI}$ | $K_{PD}$ | $\lambda$ | $\mu$ |
| Fuzzy PID | Uniform | 4.35 | 0.05 | 0.02 | 11.76 | 22.04 | - | - |
| | Logistic | 4.31 | 0.07 | 0.02 | 6.08 | 18.61 | - | - |
| | Henon | 4.28 | 0.06 | 0.02 | 7.64 | 23.45 | - | - |
| Fuzzy FOPID | Uniform | 4.56 | 0.72 | 0.47 | 1.28 | 2.07 | 0.87 | 0.87 |
| | Logistic | 4.37 | 0.15 | 0.18 | 6.29 | 3.53 | 0.78 | 0.99 |
| | Henon | 4.25 | 0.22 | 0.25 | 3.17 | 4.00 | 0.99 | 0.84 |

It is already mentioned that for the case of the PID controller, the same value of $J_{min}$ is obtained for all the three different versions of PSO. However from Table 2, it is observed that the best $J_{min}$ is obtained by the Henon map adapted PSO algorithm for the fuzzy PID and the fuzzy FOPID controller. Also the convergence characteristics for these two cases as shown in Figure 6 indicate that the Henon map adapted PSO converges to the solution quickly as compared to the Logistic map version. The results indicate towards a general trend that the chaotic map adapted PSO algorithms are better in those cases where the number of parameters to be tuned are more and the objective function is obtained through nonlinear relationships with long memory (for fuzzy logic and fractional calculus respectively). For PID controller design, there is no difference in the minima found by three PSO variants but Logistic map assisted version of PSO takes less iterations to converge (as can be observed from Figure 6). For Fuzzy PID/FOPID design, both the minima of the control objective and number of iterations is lower with Henon map augmented PSO than that with the other variants.

The controller design task has been carried out with the scenario shown in Figure 2 for multiple realizations of the stochastic processes in the generated and demand powers. Both the wind and solar power have superimposed fluctuations about the steady state value. Both the powers drop to significantly different levels after 40 sec. This is representative of the actual scenario where there is a high variability in the generated power over time, depending on the weather condition. The load demand also has similar fluctuations about the steady state values and increases suddenly after 80 sec. The controller tuning methodology takes these fluctuations into account with multiple such realizations of the stochastic fluctuations while computing the controller gains. Therefore, the controller is expected to work in a wide variety of scenarios.

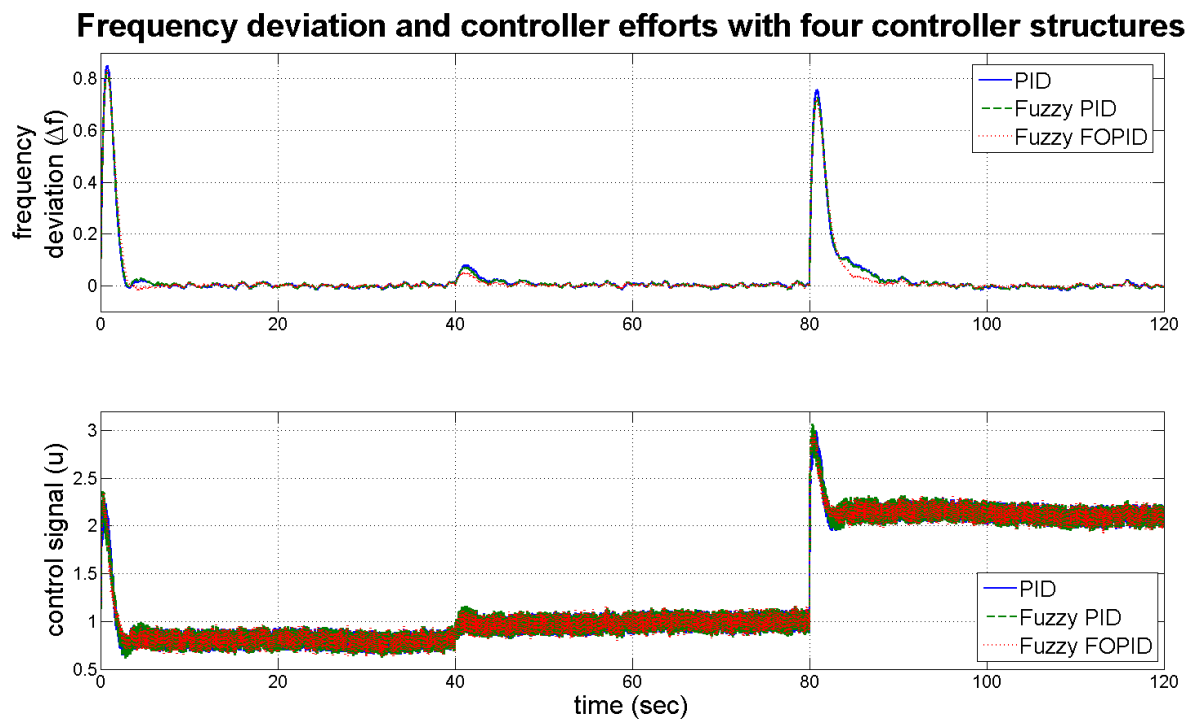

Figure 7: Deviation in frequency and control signal with different controller structures.

Figure 7 shows the frequency deviation for the three different controller structures with the best obtained values in Table 2. The corresponding individual powers of the different components of the hybrid power system for these cases are shown in Figure 8. From the frequency deviation curves, it is difficult to distinguish that the fuzzy FOPID controller works better than the other two, but this is indicated by the numerical values of $J_{min}$ in Table 2. However, from the control signal curve in Figure 7, it can be observed that the band of oscillations for the fuzzy FOPID controller is less than that with the PID or the fuzzy PID controller. This is especially important from the practical implementation point of view because the control signal actuates mechanical components like the FESS, BESS, DEG etc.

Sustained oscillation in the actuator command would wear out the mechanical components and would deteriorate the life time and the performance of that particular component. Figure 8 also shows that amongst different energy storing or supplying components, the UC contributes to the maximum power followed by the FESS, BESS, DEG and FC respectively. Also, in Figure 8 positive powers in FC, DEG indicate that they are power producing and conversely negative powers in FESS, BESS, UC signify that they are energy storing elements in the hybrid power system.

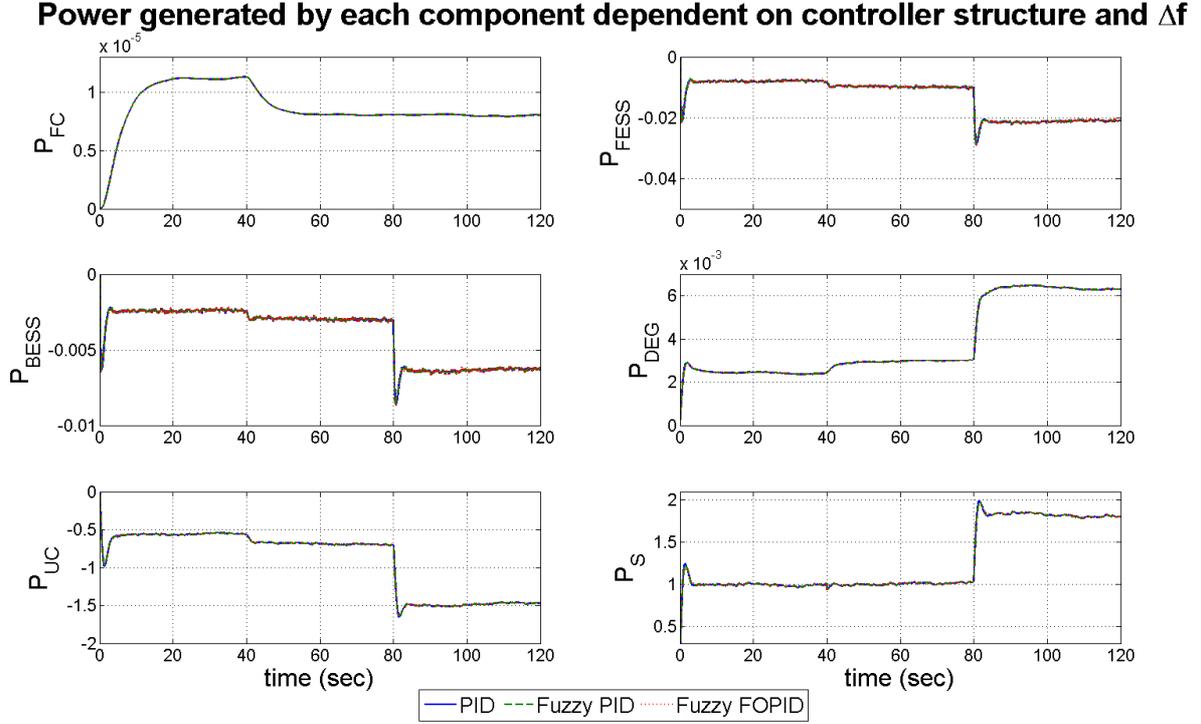

Figure 8: Power generated by different components of the hybrid power system.

TABLE 3: ROBUSTNESS TEST AGAINST PARAMETER VARIATION OF UC

| Condition | Performance - ISE (ISDCO) for different controllers | | |
|---|---|---|---|
| | PID | Fuzzy PID | Fuzzy FOPID |
| nominal | 1.55 (2.96) | 1.50 (2.78) | 1.50 (2.76) |
| 30% increase | 1.24 (13.67) | 1.18 (13.53) | 1.17 (13.47) |
| 50% increase | 1.14 (26.71) | 1.07 (26.59) | 1.06 (26.51) |
| 30% decrease | 2.37 (49.09) | 2.4 (48.85) | 2.36 (48.87) |
| 50% decrease | 3.89 (236.7) | 4.17 (236.3) | 4.00 (236.4) |

To test the robustness of the obtained solutions and compare them across different controllers, two kinds of simulations are shown next. In section 5.2, the parameters of the transfer function of the maximum power storing/producing component (i.e. UC) are changed and its effect on the system performance is studied. In section 5.3, different components are disconnected from the hybrid power system one by one and the performance of the controller to adjust the dynamics of the remaining power system is quantified by the increase in $J$ (15).

## 5.2. *Robustness against ultracapacitor parameter variation*

A 30% and 50% increase and decrease of UC gain and time constant has been introduced to test the robustness of three controller structures. Among all the components connected in the feedback path, the UC has the highest share of power as can be seen from Figure 8. Hence changes in the UC parameters will affect the overall system more than the other components. So testing for parametric robustness of UC is essentially testing for the worst case scenario. Figure 9 shows the frequency and control deviation for the three controller structures. Table 3 lists the corresponding values of the performance measures (ISE and ISDCO) for these perturbed cases of UC parameters. The robustness of the fuzzy FOPID controller as the centralized controller in feedback loop is evident since it consistently keeps ISE and ISDCO at lower values than that with the PID and fuzzy PID structures.

From Table 3, it is evident that the fuzzy FOPID outperforms the PID or the fuzzy PID controller in all the perturbed cases although for the nominal case the performance gain is relatively smaller. The improvement with fuzzy FOPID over fuzzy PID is small but still the former is better under all cases of (±) parameter perturbation of UC.

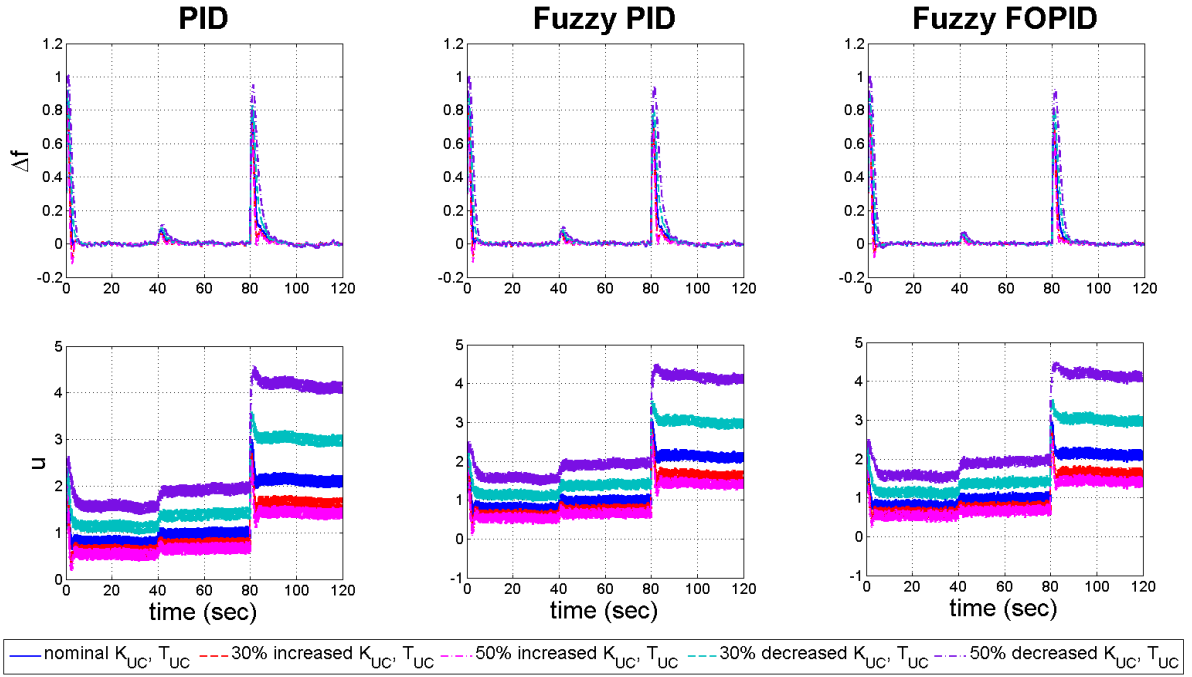

Figure 9: Frequency and control signal deviations with variation UC parameter.

## 5.3. *Robustness against disconnecting different energy storage components*

Robustness of the obtained solutions are verified next by disconnecting different components and looking at the two performance measures i.e. ISE/ISDCO and their percentage change from the nominal case (having all the components). Three cases are considered i.e. separately disconnecting the DEG, FESS and BESS.

The corresponding performance measures ($I$) is given by (23) along with the decrease in performance due to disconnecting a particular component $k$.

$$\text{Performance decrease}_k = \left( \left( I_{\text{nominal}} - I_k \right) / I_{\text{nominal}} \right) \times 100\%,$$
$$I_{\text{nominal}} > I, I \in \{ISE, ISDCO, J\}, k \in \{DEG, FESS, BESS\}$$
(23)

From the tabulated results in Table 4, it can be seen that in all cases, the grid frequency oscillation suppression (load rejection ISE) and controller effort (ISDCO) of the PID controller is the worst. The performance is better with fuzzy PID and the best with the fuzzy FOPID controller. Table 4 shows that disconnecting FESS has higher impact on the performance followed by BESS and DEG. Also, the severity of the performance deterioration is the minimum for fuzzy FOPID, followed by fuzzy PID and the traditional PID controller.

TABLE 4: ROBUSTNESS AGAINST DISCONNECTING STORING/GENERATING ELEMENTS

| Controller | Element opened | Performance measure | | | % performance decrease | | |
|---|---|---|---|---|---|---|---|
| | | ISE | ISDCO | J | ISE | ISDCO | J |
| PID | nominal | 1.55 | 2.96 | 4.51 | - | - | - |
| | DEG | 1.55 | 3.01 | 4.56 | 3.54 | 9.29 | 7.26 |
| | FESS | 1.61 | 3.29 | 4.90 | 7.20 | 19.56 | 15.21 |
| | BESS | 1.56 | 3.05 | 4.61 | 4.27 | 10.74 | 8.46 |
| Fuzzy PID | nominal | 1.50 | 2.78 | 4.28 | - | - | - |
| | DEG | 1.51 | 2.82 | 4.33 | 0.60 | 2.50 | 1.83 |
| | FESS | 1.57 | 3.10 | 4.67 | 4.47 | 12.52 | 9.69 |
| | BESS | 1.52 | 2.86 | 4.38 | 1.33 | 3.92 | 3.01 |
| Fuzzy FOPID | nominal | 1.50 | 2.76 | 4.25 | - | - | - |
| | DEG | 1.51 | 2.80 | 4.31 | 0.60 | 1.63 | 1.27 |
| | FESS | 1.57 | 3.07 | 4.64 | 4.94 | 11.36 | 9.10 |
| | BESS | 1.52 | 2.84 | 4.36 | 1.40 | 3.01 | 2.44 |

## 6. Effect of nonlinear operation of the energy storing/producing elements in the feedback path

In order to test the robustness of the proposed control scheme for significant nonlinearity in the energy storing or generating devices like FESS, BESS, UC and DEG, all the components are considered to have a rate constraint type nonlinearity. This typical nonlinearity restricts a particular component to store or release power very fast by putting an upper and lower bound on them which is representative of a realistic scenario. Figure 10 shows the implementation of a rate constraint non-linearity for all the different types of energy generation and storage subsystems which are represented by first order transfer functions. It is implemented as a saturation block (with pre-specified upper and lower cut-off limits) before the integrator block.

Figure 11 shows this phenomenon for each of the four components in feedback path with a constraint of $|\dot{P}_{FESS}|<0.02$, $|\dot{P}_{BESS}|<0.005$, $|\dot{P}_{UC}|<1.2$, $|\dot{P}_{DEG}|<0.001$, while the controllers are tuned for the linear operation, as considered in the previous sections. The robustness comparison can be done with the increase in the cost function (15) for the three controller structures as reported in Figure 11 which clearly shows that fuzzy FOPID outperforms the other two structures. Figure 12 also shows the deviation between the

corresponding linear operation and nonlinear rate constrained operation with the well-tuned controllers. The value of the objective function reported alongside each controller structure in Figure 12 again verifies the fact that the fuzzy FOPID controller is better than its counterparts.

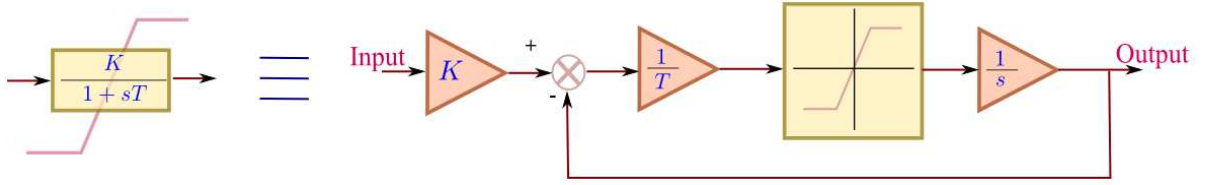

Figure 10: Rate constraint type nonlinearity for a first order transfer function representing the generation or storage units in the feedback path.

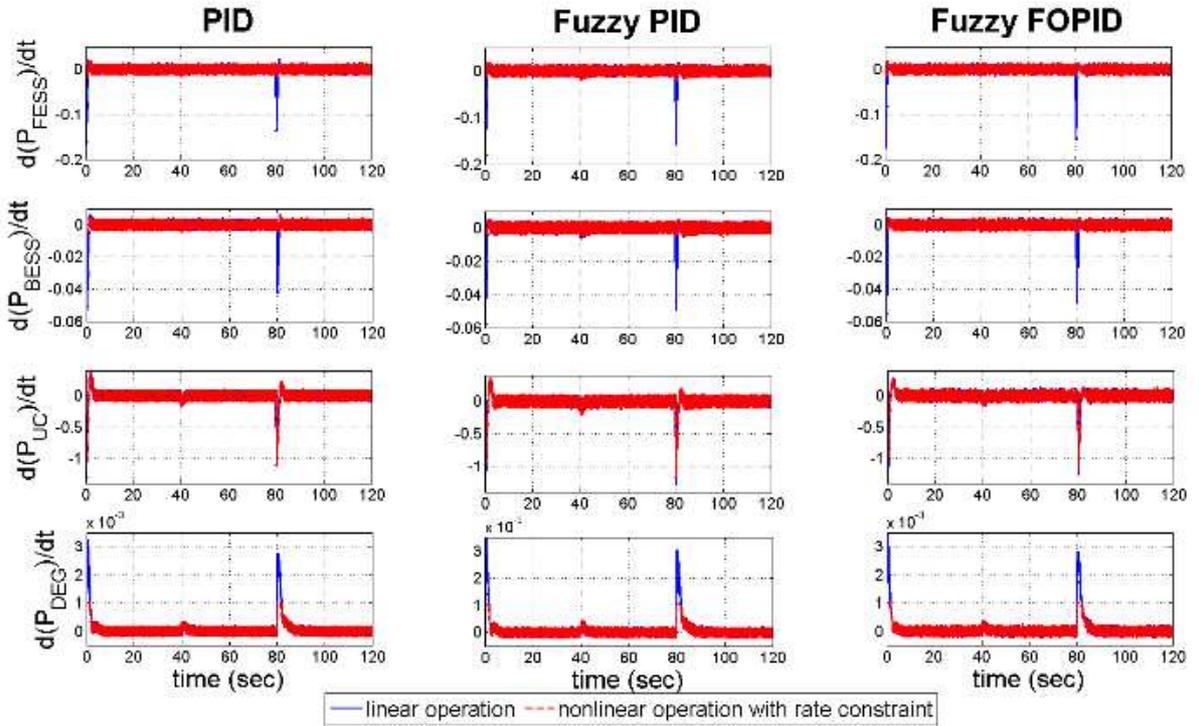

Figure 11: Effect of nonlinear rate constraints on energy storage/generating elements.

## 7. Discussion

We have initially tuned the three controllers with linear models of the hybrid power system components. Then, we have tested the performance of the controllers under nonlinear operation of the energy storing and producing elements in section 6. In this mode of operation, all the power system components have a rate constraint type nonlinearity which is representative of a more realistic scenario [20]. Since our optimization methodology is generic in nature (unlike $H_2/H_\infty$ techniques [17] or Linear Matrix Inequality (LMI) based techniques [33] etc.) and the controllers are not limited to have a linear structure only, it is easy to explore other kinds of process nonlinearities as well. The fractional order fuzzy controllers also have high robustness properties as demonstrated in sections 5.2 and 5.3. This is important since the control design can take into account the effects of different un-modelled dynamics which are neglected while modelling the different power system components. Therefore in spite of the consideration of small signal linear models in the initial power system components (i.e. during the tuning phase), the other advantages of robustness e.g. parametric uncertainties, robustness against disconnection of different energy storage components, satisfactory performance in the presence of rate constraint nonlinearities (as in section 6) are also enjoyed due to the presence of fuzzy logic in the FO controller. Therefore, it might be an attractive option to the system designer for giving preference to the fuzzy control design over other traditional techniques.

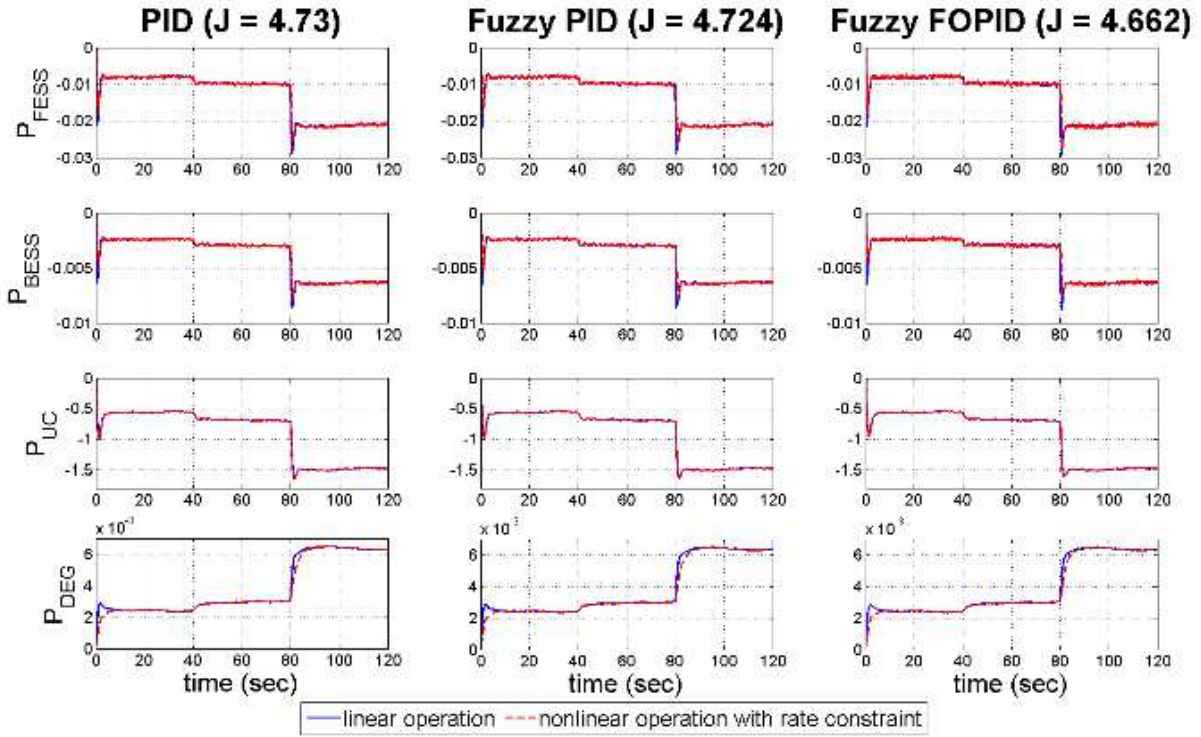

Figure 12: Change in performance of the storage elements due to nonlinear rate constraints.

Also, here other controller design approaches like model predictive control, sliding mode control etc. could have been used which might give better performance over the PID controller for this hybrid power system. However, most industrial control systems still rely on the PID controller due to its simple structure, ease of implementation and satisfactory closed loop performance in the presence of uncertainty. Therefore the objective of introducing the simulation results with the PID controller is solely to serve as a benchmark of traditionally accepted industrial practices. The focus of the present study is to make a comparison between the integer order fuzzy PID controller and the fractional order fuzzy PID controller. Naturally these are more complicated control system designs and would require more expensive hardware to implement. The system designer can look at the relative improvement in performance as obtained by these fuzzy strategies over the simple PID controller and decide whether it is worthwhile for his specific application to obtain this improvement in control system performance, at the expense of investing in more sophisticated hardware for the fuzzy controllers.

## 8. Conclusions

This paper proposes a centralized control scheme with a novel fractional order fuzzy PID controller for suppressing the grid frequency oscillation in a hybrid power system. The centralized scheme offers the advantage of cost effectiveness, reduced maintenance, wiring and number of parameters to tune. Parameters of the fuzzy FOPID controller are tuned with chaotic map adapted PSO algorithms and the controller outperforms the PID and fuzzy PID controller structures. The chaotic map adapted PSO works better than the traditional PSO in terms of the quality of the solution and obtaining faster convergence. The fuzzy FOPID controller also shows high robustness properties with respect to parameter variation in UC, nonlinear rate constraint on feedback elements and also on disconnection of some components. This suggests that once the fuzzy FOPID is tuned for the nominal system and implemented, it would not need additional retuning or online auto-tuning, even in the perturbed cases, which increases system reliability. The operation of each component in the nonlinear regime is also probed into and it is shown that the fuzzy FOPID controller is able to handle system nonlinearities better than the other two controller structures.